\documentclass[12pt]{article}

\usepackage[centertags]{amsmath}
\usepackage{amsfonts}
\usepackage{amssymb}
\usepackage{amsthm}
\usepackage{newlfont}
\usepackage{epsfig}
\usepackage{amscd}

\newcommand{\NN}{{\mathbb N}}

\newcommand{\RR}{{\mathbb R}}

\newcommand{\CC}{{\mathbb C}}
\newcommand{\beq}{\begin{equation}}
\newcommand{\eeq}{\end{equation}}
\newcommand{\ba}{\begin{array}}
\newcommand{\ea}{\end{array}}
\newcommand{\bea}{\begin{eqnarray}}
\newcommand{\eea}{\end{eqnarray}}

\begin{document}

\begin{center}
{\large \sc \bf On the solutions of the second heavenly and \\ Pavlov equations}
\vskip 20pt

{\large  S. V. Manakov$^{1,\S}$ and P. M. Santini$^{2,\S}$}

\vskip 20pt

{\it 
$^1$ Landau Institute for Theoretical Physics, Moscow, Russia

\smallskip

$^2$ Dipartimento di Fisica, Universit\`a di Roma "La Sapienza", and \\
Istituto Nazionale di Fisica Nucleare, Sezione di Roma 1 \\
Piazz.le Aldo Moro 2, I-00185 Roma, Italy}

\bigskip

$^{\S}$e-mail:  {\tt manakov@itp.ac.ru, paolo.santini@roma1.infn.it}

\bigskip

{\today}

\end{center}

\begin{abstract}
We have recently solved the inverse scattering problem for one parameter families 
of vector fields, and used this result to construct the formal solution of the Cauchy 
problem for a class of integrable nonlinear partial differential equations  
connected with the commutation of multidimensional vector 
fields, like the heavenly equation of Plebanski, the dispersionless Kadomtsev 
- Petviashvili (dKP) equation and the two-dimensional dispersionless Toda (2ddT) equation, as well 
as  with the commutation of one 
dimensional vector fields, like the Pavlov equation. We also showed that 
the associated Riemann-Hilbert inverse problems are powerfull tools to establish if the solutions 
of the Cauchy problem break at finite time,    
to construct their longtime behaviour and characterize classes of implicit solutions.   
In this paper, using the above theory, we concentrate on the heavenly and Pavlov equations,    
i) establishing that their localized solutions evolve without breaking, unlike the cases of dKP and 2ddT; 
ii) constructing the longtime behaviour of the solutions of their Cauchy problems; iii) characterizing 
a distinguished class of implicit solutions of the heavenly equation.

\newpage
 
\end{abstract}

%%%%%%%%%% %%%%%%%%%%%%%%%%%%%%%%%%%%%%%%%%%%%%%%%%%%%%%%%
\section{Introduction}
%%%%%%%%%%%%%%%%%%%%%%%%%%%%%%%%%%%%%%%%%%%%%%%%%%%%%%%%%%
It was observed long ago \cite{ZS} that the commutation of multidimensional 
vector fields can generate integrable nonlinear partial differential equations (PDEs) in arbitrary 
dimensions. Some of these equations are dispersionless (or quasi-classical) limits 
of integrable PDEs, having the dispersionless Kadomtsev 
- Petviashvili (dKP) equation \cite{Timman},\cite{ZK} as universal prototype example; 
they arise in various problems of 
Mathematical Physics and are intensively studied in the recent literature 
(see, f.i., \cite{KG} - \cite{FM}). In particular, an elegant integration scheme applicable, in general,  
to nonlinear PDEs associated with Hamiltonian vector fields, was presented in \cite{Kri} and a 
nonlinear $\bar\partial$ - dressing was developed in \cite{K-MA-R}. Special classes of nontrivial solutions 
were also derived (see, f.i., \cite{DT}, \cite{G-M-MA}).

Distinguished examples of PDEs arising as the commutation conditions $[\hat L_1(\lambda),\hat L_2(\lambda)]=0$ of pairs of one 
parameter families of vector fields, being $\lambda\in\CC$ the spectral parameter, are the following. \\
1. The vector nonlinear PDE in $N+4$ dimensions \cite{MS1}: 
\beq
\label{quasilin-U}
\vec U_{t_1z_2}-\vec U_{t_2z_1}+
\left(\vec U_{z_1}\cdot\nabla_{\vec x}\right)\vec U_{z_2}-\left(\vec U_{z_2}\cdot\nabla_{\vec x}\right)\vec U_{z_1}=\vec 0,
\eeq
where $\vec U(t_1,t_2,z_1,z_2,\vec x)\in\RR^N$, $\vec x=(x^1,\dots,x^N)\in\RR^N$ and  
$\nabla_{\vec x}=(\partial_{x_1},..,\partial_{x_N})$, associated with the following pair of $(N+1)$ dimensional 
vector fields
\beq
\label{L1L2_quasilin-U}
\hat L_i=\partial_{t_i}+\lambda\partial_{z_i}+\vec U_{z_i}\cdot\nabla_{\vec x},~~~~i=1,2. 
\eeq  
2. Its dimensional reduction, for $N=2$ \cite{MS1}:
\beq
\label{quasilin-U-bis}
\ba{l}
\vec U_{tx}-\vec U_{zy}+\left(\vec U_{y}\cdot\nabla_{\vec x}\right)\vec U_{x}-
\left(\vec U_{x}\cdot\nabla_{\vec x}\right)\vec U_{y}=\vec 0, \\
\vec U\in\RR^2,~~\vec x=(x,y),~~\nabla_{\vec x}=(\partial_x,\partial_y),
\ea
\eeq
obtained renaming the independent variables as follows: $t_1=z,~t_2=t,~~x_1=x,~~x_2=y$, associated with the 
two-dimensional vector fields:
\beq
\label{L1L2_quasilin-U-bis}
\ba{l}
\hat L_1=\partial_{z}+\lambda\partial_{x}+\vec U_x\cdot\nabla_{\vec x},     \\
\hat L_2=\partial_{t}+\lambda\partial_{y}+\vec U_y\cdot\nabla_{\vec x}.
\ea
\eeq
3. The divergenceless reduction $\nabla_{\vec x}\cdot\vec U=0$ of (\ref{quasilin-U-bis}), the celebrated second 
heavenly equation of Plebanski \cite{Pleb}: 
\beq
\label{heavenly}
\theta_{tx}-\theta_{zy}+\theta_{xx}\theta_{yy}-\theta^2_{xy}=0,~~~
\theta=\theta(x,y,z,t)\in\RR,~~~~~x,y,z,t\in\RR,
\eeq
describing self-dual vacuum solutions of the Einstein equations, associated with 
the following pair of Hamiltonian two-dimensional vector fields  
\beq
\label{L1L2_heav}
\ba{l}
\hat L_1\equiv \partial_{z}+\lambda\partial_{x}+\theta_{xy}\partial_{x}-\theta_{xx}\partial_{y}, \\
\hat L_2\equiv \partial_{t}+\lambda\partial_{y}+\theta_{yy}\partial_{x}-\theta_{xy}\partial_{y}. 
\ea
\eeq
4. The following system of two nonlinear PDEs in $2+1$ dimensions \cite{MS3}:
\beq
\label{dKP-system}
\ba{l}
u_{xt}+u_{yy}+(uu_x)_x+v_xu_{xy}-v_yu_{xx}=0,         \\
v_{xt}+v_{yy}+uv_{xx}+v_xv_{xy}-v_yv_{xx}=0,
\ea
\eeq
arising from the commutation of the two-dimensional vector fields \cite{MS3} 
\beq\label{Ltilde}
\ba{l}
\tilde L_1\equiv \partial_y+(\lambda+v_x)\partial_x-u_x\partial_{\lambda}, \\
\tilde L_2\equiv \partial_t+(\lambda^2+\lambda v_x+u-v_y)\partial_x+(-\lambda u_x+u_y)\partial_{\lambda}, 
\ea
\eeq
and describing a general integrable Einstein-Weyl metric \cite{Dunajj}.                       \\
5. The $v=0$ reduction of (\ref{dKP-system}), the celebrated dKP equation  
\beq
\label{dKP}
(u_t+uu_x)_x+u_{yy}=0,~~~
u=u(x,y,t)\in\RR,~~~~~x,y,t\in\RR,
\eeq 
(the $x$-dispersionless limit of the celebrated Kadomtsev-Petviashvili equation \cite{KP}),  
associated with the following pair of Hamiltonian two-dimensional vector fields \cite{Zakharov,Kri}:
\beq
\label{L1L2_dKP}
\ba{l}
\hat L_1\equiv \partial_y+\lambda\partial_x-u_x\partial_{\lambda},    \\
\hat L_2\equiv \partial_t+(\lambda^2+u)\partial_x+(-\lambda u_x+u_y)\partial_{\lambda},
\ea
\eeq
describing the evolution of small amplitude, nearly one-dimensional waves in shallow water \cite{AC} 
near the shore (when the $x$-dispersion can be neglected), as well as unsteady motion in transonic flow 
\cite{Timman} and nonlinear acoustics of confined beams \cite{ZK}.     \\ 
6. The $u=0$ reduction of (\ref{dKP-system}), the Pavlov equation \cite{Pavlov}
\beq 
\label{Pavlov}
\ba{l}
v_{xt}+v_{yy}=v_yv_{xx}-v_xv_{xy},~~~~~~~ 
v=v(x,y,t)\in\RR,~~~~~x,y,t\in\RR,
\ea
\eeq
associated with the non-Hamiltonian one-dimensional vector fields \cite{Duna}
\beq
\label{L1L2_Pavlov}
\ba{l}
\hat L_1\equiv \partial_y+(\lambda +v_x)\partial_x, \\
\hat L_2\equiv \partial_t+(\lambda^2+\lambda v_x-v_y)\partial_x.
\ea
\eeq
7. The two-dimensional dispersionless Toda (2ddT) equation \cite{FP,Zak}:
\beq\label{2ddT}
\phi_{\zeta_1\zeta_2}=\left(e^{\phi_t}\right)_t,~~~\phi=\phi(\zeta_1,\zeta_2,t)   
\eeq
(or $\varphi_{\zeta_1\zeta_2}=\left(e^{\varphi}\right)_{tt},~\varphi=\phi_t$), associated with the pair of 
Hamiltonian vector fields \cite{TT1}:    
\beq\label{L1L2_2ddT}
\ba{l}
\hat L_1=\partial_{\zeta_1}+\lambda e^{\frac{\phi_{t}}{2}}\partial_{t}+
\left(-\lambda (e^{\frac{\phi_{t}}{2}})_t+\frac{\phi_{\zeta_1 t}}{2}\right)\lambda\partial_{\lambda} ,\\ 
\hat L_2=\partial_{\zeta_2}+\lambda^{-1} e^{\frac{\phi_{t}}{2}}\partial_{t}+
\left(\lambda^{-1}(e^{\frac{\phi_{t}}{2}})_t-\frac{\phi_{\zeta_2 t}}{2}\right)\lambda\partial_{\lambda}, 
\ea
\eeq  
describing integrable heavens \cite{BF,GD} and Einstein - Weyl geometries \cite{H}, \cite{J}, \cite{Ward};     
whose string equations solutions \cite{TT2} are relevant in the ideal Hele-Shaw problem 
\cite{MWZ}-\cite{MAM}.

The Inverse Spectral Transform (IST) for $1$-parameter families of multidimensional vector fields,  
developed in \cite{MS1} (see also \cite{MS2}),  
has allowed one to construct the formal solution of the Cauchy problem for the nonlinear PDEs 
(\ref{quasilin-U-bis}) and (\ref{heavenly}) in \cite{MS1}, for equations (\ref{dKP-system}) and (\ref{dKP}) 
in \cite{MS3}, for equation (\ref{Pavlov}) in \cite{MS4} and for the wave form 
$(e^{\phi_t})_t=\phi_{xx}+\phi_{yy}$ of equation (\ref{2ddT}) in \cite{MS5}.    
This IST, introducing interesting novelties 
with respect to the classical IST for soliton equations \cite{ZMNP,AC}, turns out to be, together with its associated
 nonlinear Riemann - Hilbert (RH) dressing, an efficient 
tool to study several properties of the solution space of the PDE under consideration: 
i) the characterization of a 
distinguished class of spectral data for which the associated nonlinear RH problem is linearized and solved, 
corresponding to a class of implicit solutions of the PDE (for the dKP and 2ddT equations respectively 
in \cite{MS6} and in \cite{MS5}, and for the Dunajski generalization \cite{Duna} of the heavenly equation in \cite{BDM});  
ii) the construction of the longtime behaviour of the solutions of the Cauchy problem (for the dKP and 
2ddT equations respectively in \cite{MS6} and \cite{MS5});     
iii) the possibility  to establish  whether or not the lack of dispersive terms in the nonlinear PDE 
causes the breaking of localized initial profiles (for the dKP and 2ddT equations respectively in \cite{MS6} and in \cite{MS5}) 
and, if yes, to investigate in a surprisingly 
explicit way the analytic aspects of such a wave breaking, as it was done for the dKP equation  
in \cite{MS6}. 

In this paper we first develop, in \S 2, the RH dressing schemes for equations (\ref{heavenly}) 
and (\ref{Pavlov}). Then we use these two schemes \\
i) to show that, in contrast to the cases of the dKP and 2ddT equations, 
localized solutions of the PDEs (\ref{heavenly}) and (\ref{Pavlov}) do not break (in \S 2); \\   
ii) to construct the longtime behaviour of the solutions of their Cauchy problems (in \S 3); \\
iii) to characterize a 
distinguished class of spectral data for which the nonlinear RH problem of equation (\ref{heavenly}) is linearized 
and solved, corresponding to a distinguished class of implicit solutions of (\ref{heavenly}) (in \S 4).    
 
%%%%%%%%%%%%%%%%%%%%%%%%%%%%%%%%%%%%%%%%%%%%%%%%%%%%%%%%%%%%%%%%%%%%%%%%%%%%%%%%%
\section{Riemann - Hilbert dressing}
%%%%%%%%%%%%%%%%%%%%%%%%%%%%%%%%%%%%%%%%%%%%%%%%%%%%%%%%%%%%%%%%%%%%%%%%%%%%%%%%

In this section we present the RH dressing schemes for equations (\ref{heavenly}) and (\ref{Pavlov}),  
and show that localized solutions of these two models do not break. Both schemes can be extracted from 
the inverse spectral transforms presented in \cite{MS1} and in \cite{MS4}. In solving the Cauchy problems 
for equations (\ref{heavenly}) and (\ref{Pavlov}), the RH data of these dressing schemes are 
connected to the initial conditions via the direct spectral transforms presented in \cite{MS1} and in \cite{MS4}.  
 
\vskip 5pt
\noindent
{\bf 1. RH dressing for (\ref{heavenly})}. Consider the vector nonlinear RH problem on the real line:  
\beq\label{RH_heav}
\vec\pi^+(\lambda)=\vec\pi^-(\lambda)+\vec R(\vec\pi^-(\lambda),\lambda),~~\lambda\in\RR,
\eeq
where $\vec\pi^{+}(\lambda),\vec\pi^{-}(\lambda)\in\CC^2$ are $2$-dimensional 
vector functions analytic in the upper and lower 
halves of the complex $\lambda$ plane, normalized as follows 
\beq\label{normal_heav}
\ba{l}
\vec\pi^{\pm}(\lambda)=\vec\nu(\lambda;x,y,z,t)+O(\lambda^{-1}),~~|\lambda |>>1, \\
\vec\nu(\lambda;x,y,z,t)=\left(
\ba{c}
x-\lambda z \\
y-\lambda t
\ea
\right),
\ea
\eeq 
and the spectral data $\vec R(\vec \zeta,\lambda)=
(R_1(\zeta_1,\zeta_2,\lambda),R_2(\zeta_1,\zeta_2,\lambda))\in\CC^2$, defined for 
$\vec \zeta\in\CC^2,~~\lambda\in\RR$, satisfy the following properties:
\beq\label{R_heav}
\ba{ll}
\vec{\cal R}(\overline{ \vec{\cal R}(\bar{\vec\zeta},\lambda)},\lambda)=\vec\zeta,
~~~\forall\vec\zeta\in\CC^2,  &  \mbox{reality constraint,} \\
\{{\cal R}_1,{\cal R}_2\}_{\vec\zeta}=1,  &  \mbox{heavenly constraint},
\ea
\eeq
where $\vec{\cal R}(\vec\zeta,\lambda):=\vec\zeta+\vec R(\vec\zeta,\lambda)$ and $\{\cdot ,\cdot \}_{\vec\zeta}$ 
is the usual Poisson bracket with respect to the variables $\vec\zeta=(\zeta_1,\zeta_2)$. Then, assuming uniqueness of the solution 
of such a RH problem and of its linearized version, it follows that $\vec\pi^{\pm}$ are solutions of the linear problems 
$\hat L_{1,2}\vec\pi^{\pm}=\vec 0$, where $\hat L_{1,2}$ are defined in (\ref{L1L2_heav}), and   
\beq\label{pot_1_heav}
\left(
\ba{c}
\theta_y \\
-\theta_x
\ea
\right)=\vec F(x,y,z,t) \in\RR^2
\eeq
is solution of the heavenly equation (\ref{heavenly}), where
\beq\label{pot_2_heav}  
\vec F(x,y,z,t)=\int\limits_{\RR}\frac{d\lambda}{2\pi i}
\vec R\Big(\pi^-_1(\lambda;x,y,z,t),\pi^-_2(\lambda;x,y,z,t),\lambda \Big).
\eeq
As a consequence of equations $\hat L_{1,2}\vec\pi^{\pm}=\vec 0$, it follows that, for $|\lambda |>>1$: 
\beq\label{pi_large_heav}
\ba{l}
\pi^{\pm}_1=x-\lambda z-\theta_y \lambda^{-1}+\theta_t \lambda^{-2} +O(\lambda^{-3}), \\
\pi^{\pm}_2=y-\lambda t+\theta_x \lambda^{-1}-\theta_z \lambda^{-2} +O(\lambda^{-3}).
\ea
\eeq
{\it Proof}. We apply the operators $\hat L_{1,2}$ in (\ref{L1L2_quasilin-U-bis}) to the RH problem (\ref{RH_heav}), 
where
\beq\label{defU}
\vec U=-\displaystyle\lim_{\lambda\to\infty}{\lambda\left(
\vec\pi^{\pm}-
\left(
\ba{c}
x-\lambda z \\
y-\lambda t
\ea
\right) \right)},
\eeq 
obtaining the vector equation $\hat L_j\vec\pi^{+}=J(\vec\pi^{-},\lambda)\hat L_j\vec\pi^{-},~j=1,2$, where $J$ is the Jacobian matrix  
of the transformation (\ref{RH_heav}): $J_{kl}(\vec\zeta,\lambda)=\partial{\cal R_k}(\vec\zeta,\lambda)/\partial \zeta_l$. Since, 
due to (\ref{defU}), $\hat L_j\vec\pi^{\pm}\to 0$ as $\lambda\to\infty$, by uniqueness we infer that $\vec\pi^{\pm}$  
are eigenfunctions of the vector fields $\hat L_j$: $\hat L_j\vec\pi^{\pm}=\vec 0$. Consequently, $\vec U$ is solution of  
equation (\ref{quasilin-U-bis}). 
In addition, if the reality constraint (\ref{R_heav}a) is imposed, then, by uniqueness, it follows, from (\ref{RH_heav}), that 
$\overline{\vec\pi^+}=\vec\pi^-$, implying the reality of $\vec U$. At last, if the heavenly constraint (\ref{R_heav}b) 
is imposed, then $\{\pi^+_1,\pi^+_2\}=\{\pi^-_1,\pi^-_2\}$. Since 
$\{\pi^{\pm}_1,\pi^{\pm}_2\}\to 1$ as $|\lambda |\to\infty$, the analyticity of the eingenfunctions implies that 
$\{\pi^+_1,\pi^+_2\}=\{\pi^-_1,\pi^-_2\}=1$. Applying $\hat L_{1,2}$ to these two equations, one infers that 
$\nabla_{\vec x}\cdot\vec U=0$, 
implying the existence of a potential $\theta$ such that $\vec U=(\theta_y,-\theta_x)$, and implying also that 
the two vector fields $\hat L_{1,2}$ are Hamiltonian, with Hamiltonians $(H_1,H_2)=\nabla_{\vec x}\theta$,  
reducing to the vector fields (\ref{L1L2_heav}). Then the system (\ref{quasilin-U-bis}) 
reduces to the heavenly equation (\ref{heavenly}) and equation (\ref{pot_1_heav}) follows from (\ref{defU}) and 
from the integral equations characterizing the RH problem. $\Box$   

%%%%%%%%%%%
\vskip 10pt
\noindent
{\bf 2. RH dressing for (\ref{Pavlov})}. Consider the scalar nonlinear RH problem on the real line:  
\beq\label{RH_Pavlov}
\pi^+(\lambda)=\pi^-(\lambda)+R(\pi^-(\lambda),\lambda),~~\lambda\in\RR,
\eeq
where $\pi^{+}(\lambda),\pi^{-}(\lambda)\in\CC$ are scalar functions analytic in the upper and lower 
halves of the complex $\lambda$ plane, normalized as follows
\beq\label{normal_Pavlov}
\ba{l}
\pi^{\pm}(\lambda;x,y,t)=\nu(\lambda;x,y,t)+O(\lambda^{-1}),~~|\lambda |>>1, \\
\nu(\lambda;x,y,t)=-\lambda^2 t-\lambda y+x , 
\ea
\eeq 
and the spectral datum $R(\zeta,\lambda)\in\CC$ satisfies the following reality constraint:
\beq\label{R_Pavlov}
\ba{l}
{\cal R}(\overline{{\cal R}(\bar{\zeta},\lambda)},\lambda)=\zeta,
~~~\forall\zeta\in\CC,~~~\lambda\in\RR, 
\ea
\eeq
where ${\cal R}(\zeta,\lambda):=\zeta+R(\zeta,\lambda)$. Then, assuming uniqueness of the solution of such a RH problem 
and of its linearized version, it follows that $\pi^{\pm}$ are solutions of the linear problems 
$\hat L_{1,2}\pi^{\pm}=0$, where $\hat L_{1,2}$ are defined in (\ref{L1L2_Pavlov}), and
\beq\label{pot_1_Pavlov}
v=F\left(x,y,t\right)\in \RR
\eeq
is solution of equation (\ref{Pavlov}), where
\beq\label{pot_2_Pavlov}
F\left(x,y,t \right)=
\int\limits_{\RR}\frac{d\lambda}{2\pi i}R\Big({\pi^-}(\lambda;x,y,t),\lambda\Big).
\eeq
{\it Proof}. We apply the operators $\hat L_{1,2}$ in (\ref{L1L2_Pavlov}) to the RH problem (\ref{RH_Pavlov}), where
\beq\label{def_v}
v(x,y,t)=-\displaystyle\lim_{\lambda\to\infty}{\lambda\left(\pi^{\pm}+\lambda^2 t+\lambda y -x\right)},
\eeq
obtaining 
$\hat L_j\pi^{+}=J(\pi^{-},\lambda)\hat L_j\pi^{-},~j=1,2$, where 
$J(\zeta,\lambda)=\partial{\cal R}(\zeta,\lambda)/\partial\zeta$. Since $\hat L_1\pi^{\pm}\to 0$ as $\lambda\to\infty$, 
it follows that, by uniqueness, $\pi^{\pm}$ are eigenfunctions of $\hat L_1$: $\hat L_1\pi^{\pm}=0$. Consequently, we 
have the asymptotics
\beq
\ba{l}
\pi^{\pm}=-\lambda^2 t-\lambda y +x-\frac{v}{\lambda}+\frac{\beta}{\lambda^2}+O(\lambda^{-3}),~~|\lambda |>>1. \\
\beta_x=v_y+v^2_x ,
\ea
\eeq 
implying that also $\hat L_2\pi^{\pm}\to 0$ as $\lambda\to\infty$. Therefore, again by uniqueness, $\hat L_{1,2}\pi^{\pm}=0$,  
$v$ satisfies equation (\ref{Pavlov}) and equation (\ref{pot_1_Pavlov}) follows from (\ref{def_v}) and 
from the integral equations characterizing the RH problem. If, in addition, the reality constraint (\ref{R_Pavlov}) 
is imposed, then, by uniqueness, $\overline{\pi^+}=\pi^-$ and, consequently, $v\in\RR$. $\Box$ 
 
We recall that, in the dressing schemes for the dKP and 2ddT equations, the spectral mechanism responsible for breaking is 
the presence, in the normalizations of the solutions of their RH problems, of the unknown fields, so that the inverse 
formulae define the solutions of the PDEs only implicitely \cite{MS6,MS5}. Since the normalizations (\ref{normal_heav}), 
(\ref{normal_Pavlov}) depend only on the space-time independent variables, such a breaking mechanism is absent here, 
and we expect that localized solutions of (\ref{heavenly}) and (\ref{Pavlov}) do not break during their evolution.

%%%%%%%%%%%%%%%%%%%%%%%%%%%%%%%%%%%%%%%%%%%%%%%%%%%%%%%%%%
%%%%%%%%%%%%%%%%%%%%%%%%%%%%%%%%%%%%%%%%%%%%%%%%%%%%%%%%%
\section{Longtime behaviour of the solutions}
%%%%%%%%%%%%%%%%%%%%%%%%%%%%%%%%%%%%%%%%%%%%%%%%%%%%%%%%%%

The IST provides an effective tool to construct the longtime 
behaviour of the solutions of integrable PDEs. In this section we investigate the longtime 
behaviour of the solutions of the nonlinear RH problems (\ref{RH_heav}), (\ref{RH_Pavlov}), and, consequently, 
of the solutions of the Cauchy problems for equations (\ref{heavenly}) and (\ref{Pavlov}). 

%%%%%%%%%%%%%%%%%%%%%%%%%%%%%%%%%%%%%%%%%%%%%%%%%%
\vskip 5pt
\noindent
{\bf Asymptotics of equation (\ref{heavenly})}. We study the longtime $t>>1$ regime in the space regions
\beq\label{asympt_reg_heav}
x=\tilde x+v_1t,~y=v_2t,~z=v_3t,~\tilde x,v_1,v_2,v_3=O(1),~v_2,v_3\ne 0,~t>>1.
\eeq

The system of nonlinear integral equations characterizing the solutions of the RH problem (\ref{RH_heav}) 
is conveniently written in the form:
\beq\label{phi_equ_heav}
\ba{l}
\phi_j(\lambda)= \frac{1}{2\pi i}\int_{\RR}\frac{d\lambda'}{\lambda'-(\lambda -i0)}
R_j\Big(\tilde x+(v_1-\lambda'v_3)t+\phi_1(\lambda'), \\
(v_2-\lambda')t+\phi_2(\lambda'),\lambda' \Big),~~~j=1,2, 
\ea
\eeq 
where
\beq\label{def_phi_heav}
\phi_1(\lambda)=\pi^-_1(\lambda)-(x-\lambda z),~~~\phi_2(\lambda)=\pi^-_2(\lambda)-(y-\lambda t).
\eeq
The fast decay, in $t$, of $\phi_j,~j=1,2$, due to the two 
different linear growths in $t$ of the first two 
arguments of $R_1,R_2$, is partially contrasted if $v_1=v_2v_3$; i.e., on the saddle surface:
\beq\label{saddle}
x=\tilde x+\frac{yz}{t}~~~~(v_1=v_2v_3).
\eeq 
Indeed, on such two-dimensional manifold, the first two arguments of $R_1,R_2$ in (\ref{phi_equ_heav}) 
exhibit the same linear growth in $t$:
\beq\label{phi2_equ_heav}
\ba{l}
\phi_j(\lambda)= 
\frac{1}{2\pi i}\int_{\RR}\frac{d\lambda'}{\lambda'-(\lambda -i0)}
R_j\Big(\tilde x+v_3(v_2-\lambda')t+\phi_1(\lambda'), \\
(v_2-\lambda')t+\phi_2(\lambda'),\lambda' \Big),~~~~j=1,2.
\ea
\eeq 
Since, in this case, the main contribution to the integral occurs when $\lambda'\sim v_2$, 
we make the change of variable $\lambda'=v_2+\mu'/t$, obtaining:
\beq\label{phi3_equ_heav}
\ba{l}
\phi_j(\lambda)= 
\frac{1}{2\pi it}\int\limits_{\RR}\frac{d\mu'}{\mu'/t-(\lambda-v_2-i0)}
R_j\Big(\tilde x-v_3\mu'+\phi_1(v_2+\frac{\mu'}{t}), \\
-\mu'+\phi_2(v_2+\frac{\mu'}{t}),v_2 \Big),~~~~j=1,2.
\ea
\eeq
If $|\lambda-v_2|>>t^{-1}$, equation (\ref{phi3_equ_heav}) implies that 
$\phi_{j}(\lambda)=O(t^{-1}),~j=1,2$: 
\beq\label{phi_asympt1_heav}
\ba{l}
\phi_j(\lambda)\sim 
-\frac{1}{2\pi i(\lambda -(v_2+i0))t}\int\limits_{\RR}d\mu'
R_j\Big(\tilde x-v_3\mu'+\phi_1(v_2+\frac{\mu'}{t}), \\
-\mu'+\phi_2(v_2+\frac{\mu'}{t}),v_2 \Big),~~~~j=1,2,
\ea
\eeq
while, for $\lambda -v_2=\mu /t,~\mu=O(1)$, we have that $\phi_{j}(\lambda)=O(1),~j=1,2$:
\beq\label{phi_asympt2_heav}
\ba{l}
\phi_j(v_2+\frac{\mu}{t})\sim 
\frac{1}{2\pi i}\int\limits_{\RR}\frac{d\mu'}{\mu'-(\mu-i0)}
R_j\Big(\tilde x-v_3\mu'+\phi_1(v_2+\frac{\mu'}{t}), \\
-\mu'+\phi_2(v_2+\frac{\mu'}{t}),v_2 \Big),~~~~j=1,2. 
\ea
\eeq
Therefore it is not possible to neglect, in the integral equations (\ref{phi3_equ_heav}), 
$\phi_{1,2}$ in the arguments of $R_{1,2}$ and these integral equations remain nonlinear also 
in the longtime regime. Summarizing, due to equations (\ref{phi_asympt2_heav}), (\ref{pot_1_heav}) and (\ref{pot_2_heav}), 
the longtime behaviour of the solutions of the Cauchy problem for the heavenly equation on the saddle (\ref{saddle}) 
reads:
\beq
\label{asympt_1_heav}
\ba{l}
\left(
\ba{c}
-\theta_y \\
\theta_x
\ea
\right)= \frac{1}{t}\vec F_{\infty}\left(x-\frac{yz}{t},\frac{y}{t},\frac{z}{t}\right)+o(t^{-1}),
~~x-\frac{yz}{t}=O(1),~~t>>1, 
\ea
\eeq
where
\beq\label{asympt_2_heav}
\ba{l}
\vec F_{\infty}(\tilde x,v_2,v_3)=  \\
-\frac{1}{2\pi i}\int\limits_{\RR}d\mu
\vec R\Big(\tilde x-v_3\mu+a_1(\mu;\tilde x,v_2,v_3),-\mu+a_2(\mu;\tilde x,v_2,v_3),v_2 \Big).
\ea
\eeq
and $a_j(\mu;\tilde x,v_2,v_3),~j=1,2$ are the solutions of the following nonlinear integral equations 
\beq\label{a_equ_heav}
\ba{l}
a_j(\mu)= 
\frac{1}{2\pi i}\int\limits_{\RR}\frac{d\mu'}{\mu'-(\mu-i0)}
R_j\Big(\tilde x-v_3\mu'+a_1(\mu'),-\mu'+a_2(\mu'),v_2 \Big),~j=1,2.
\ea
\eeq
Outside the saddle, the solution decays faster than $1/t$. 

We remark that the longtime behaviour (\ref{asympt_1_heav}) is formally the same as that of 
solutions of the linearized heavenly equation $\theta_{tx}-\theta_{zy}=0$; the nonlinearity manifests 
only in the fact that $\vec F_{\infty}$ is obtained, from the scattering data, through the solution 
of the nonlinear integral equations (\ref{a_equ_heav}). These equations characterize the new RH problem:   
\beq\label{RH_asympt_heav}
\vec A^+(\mu)=\vec A^-(\mu)+\vec R(\vec A^-(\mu),v_2),~~\mu\in\RR,
\eeq
where $\vec A^{\pm}(\mu;\tilde x,v_2,v_3)\in\CC^2$ are $2$-dimensional vector functions 
analytic in the upper and lower halves of the complex $\mu$ plane, normalized as follows:
\beq\label{A_asympt_heav}
\ba{l}
\vec A^{\pm}(\mu)=\left(
\ba{c}
\tilde x-v_3\mu \\
-\mu
\ea
\right)
+O(\mu^{-1}),
\ea
\eeq 
$\vec R(\vec \zeta,\lambda)$ are the heavenly spectral data and 
$\vec a(\mu;\tilde x,v_2,v_3)\equiv \vec A^-(\mu;\tilde x,v_2,v_3)-(\tilde x-v_3\mu,-\mu)^T$.   

The above analysis extends, with minor modifications, to equation (\ref{Pavlov}), for which we 
give just the results.
%%%%%%%%%%%%%%%%%%%%%%%%%%%%%%%%%%%%%%%%%%%
\vskip 5pt
\noindent
{\bf Asymptotics of equation (\ref{Pavlov})}. Let $t>>1$ and 
\beq\label{asympt_reg_Pavlov}
x=\tilde x+v_1t,~~y=v_2t,~~~\tilde x,v_1,v_2=O(1),~~v_2\ne 0,~~t>>1.
\eeq  
On the parabola 
\beq\label{parabola}
x=\tilde x-\frac{y^2}{4t}~~~(v_1=-\frac{v^2_2}{4}),
\eeq
the analogue of the heavenly saddle (\ref{saddle}), the longtime behaviour of the solutions  
of equation (\ref{Pavlov}) is given by
\beq\label{asympt_1_Pavlov}
\ba{l}
v= \frac{1}{\sqrt{t}}F_{\infty}\left(x+\frac{y^2}{4t},\frac{y}{t}\right)+
o\left(\frac{1}{\sqrt{t}}\right),
\ea
\eeq
where
\beq\label{asympt_2_Pavlov}
\ba{l}
F_{\infty}\left(\tilde x,v_2 \right)=
\frac{1}{2\pi i}\int\limits_{\RR}d\mu
R\Big(\tilde x-\mu^2+a(\mu;\tilde x,v_2),-\frac{v_2}{2} \Big),
\ea
\eeq
$R(\zeta,\lambda)$ is the spectral datum of equation (\ref{Pavlov}) , and 
$a(\mu;\tilde x,v_2)$ is the unique solution of the nonlinear integral equation
\beq\label{a_Pavlov}
\ba{l}
a(\mu)= 
\frac{1}{2\pi i}\int\limits_{\RR}\frac{d\mu'}{\mu'-(\mu-i0)}
R\Big(\tilde x+{\mu'}^2+a(\mu'),\frac{v_2}{2} \Big),
\ea
\eeq
characterizing the solution of the following scalar nonlinear Riemann problem 
on the real axis: 
\beq\label{RH_asympt_Pavlov}
\ba{l}
A^+(\mu)=A^-(\mu)+ 
R(A^-(\mu),-\frac{v_2}{2}),~~\mu\in\RR, \\
A^{\pm}(\mu;\tilde x,\frac{v_2}{2})=\tilde x - \mu^2+O(\mu^{-1}),~~|\mu|>>1,
\ea
\eeq
where $a(\mu;\tilde x,\frac{v_2}{2})\equiv A^-(\mu)-(\tilde x - \mu^2)$. 
Outside of the parabola (\ref{parabola}), the solution decays faster than $1/\sqrt{t}$.

%%%%%%%%%%%%%%%%%%%%%%%%%%%%%%%%%%%%%%%%%%%%%%%%%%%%%%%%%
\section{Implicit solutions of the heavenly equation}
%%%%%%%%%%%%%%%%%%%%%%%%%%%%%%%%%%%%%%%%%%%%%%%%%%%%%%%%%%

In this section we construct a class of explicit solutions of the vector nonlinear RH problem  
(\ref{RH_heav})-(\ref{R_heav}) and, correspondingly, a class of implicit solutions 
of the heavenly equation (\ref{heavenly}), parametrized by an arbitrary real function of two variables.

Suppose that the RH data in (\ref{RH_heav}) are given by:
\beq\label{manak_R}
R_j(\zeta_1,\zeta_2,\lambda)=\zeta_j \big(e^{i(-)^{j+1}f(\zeta_1\zeta_2,\lambda)}-1\big),~~j=1,2,
\eeq
in terms of the new real spectral function $f(\zeta,\lambda)$ depending on 
$\pi^-_{1}$ and $\pi^-_{2}$ only through their product. Then the RH problem (\ref{RH_heav}) becomes   
\beq\label{RH_heav_part}
\ba{l}
\pi^+_1=\pi^-_1e^{if\left(\pi^-_1\pi^-_2,\lambda\right)},~~~\lambda\in\RR, \\
\pi^+_2=\pi^-_2e^{-if\left(\pi^-_1\pi^-_2,\lambda\right)}
\ea 
\eeq
and the following properties hold.

\vskip 10pt
\noindent
i) The reality and the heavenly constraints (\ref{R_heav}) are satisfied.

\vskip 10pt
\noindent
ii) $\pi^+_1\pi^+_2=\pi^-_1\pi^-_2$. Consequently, ($\pi^+_1\pi^+_2$) is just a 
polynomial in $\lambda$:
\beq\label{w_heav}
\pi^+_1\pi^+_2=\pi^-_1\pi^-_2=zt\lambda^2-(xt+yz)\lambda+xy-z\theta_x+t\theta_y 
\equiv w(\lambda),
\eeq
and the vector nonlinear RH problem (\ref{RH_heav}) decouples into two linear 
scalar RH problems:
\beq\label{RH_heav_linear}
\ba{l}
\pi^+_1=\pi^-_1e^{if(w(\lambda),\lambda)}, \\
\pi^+_2=\pi^-_2e^{-if(w(\lambda),\lambda)}.
\ea
\eeq

\vskip 10pt
\noindent
iii) Since, from (\ref{RH_heav_linear}), 
\beq
\pi^+_je^{i(-)^{j}f^+(\lambda)}=\pi^-_je^{i(-)^{j}f^-(\lambda)},~~j=1,2,
\eeq
where
\beq
f^{\pm}(\lambda)=\frac{1}{2\pi i}
\int_{\RR}\frac{d\lambda'}{\lambda'-(\lambda\pm i0)}f(w(\lambda'),\lambda'),
\eeq
also $\pi^+_je^{i(-)^{j}f^+(\lambda)},~j=1,2$ are polynomials in $\lambda$.  
We expand them in powers of $\lambda$, for $|\lambda |>>1$, and introduce the notation 
\beq
<\lambda^n f>=\frac{1}{2\pi}\int_{\RR}\lambda^n f(w(\lambda),\lambda)d\lambda,~~n\in\NN .
\eeq 
From the positive power expansions, it follows that
\beq
\ba{l}
\pi^+_1e^{-if^+(\lambda)}=\pi^-_1e^{-if^-(\lambda)}=x-z\lambda -z<f>, \\
\pi^+_2e^{if^+(\lambda)}=\pi^-_2e^{if^-(\lambda)}=y-t\lambda +t<f>,
\ea
\eeq
implying the following explicit solution of the RH problem (\ref{RH_heav_linear}):
\beq\label{sol_heav}
\ba{l}
\pi^{\pm}_1=
(x-\lambda z-z<f>)e^{if^{\pm}(\lambda)}, \\
\pi^{\pm}_2=
(y-\lambda t+t<f>)e^{-if^{\pm}(\lambda)}.
\ea
\eeq
From the coefficients of the $\lambda^{-1}$ term, which must be zero, one finally obtains 
the two conditions
\beq\label{implicit_heav}
\ba{l}
\theta_y=x<f>-z\left(<\lambda f>+\frac{1}{2}<f>^2\right), \\
\theta_x=y<f>-t\left(<\lambda f>-\frac{1}{2}<f>^2\right).
\ea
\eeq      
Since $<f>$ and $<\lambda f>$ depend, through $w$, on $\theta_x$ and $\theta_y$ (see (\ref{w_heav})), 
it follows that (\ref{implicit_heav}) is a system of algebraic equations defining implicitely   
the solution ($\theta_x,\theta_y$) of the heavenly equation.

In addition, equating to zero the coefficients of the $\lambda^{-2}$ terms (using (\ref{pi_large_heav})), one obtains:
\beq\label{coeff2}
\ba{l}
\theta_t=\theta_y<f>-x\left(<\lambda f>+\frac{1}{2}<f>^2 \right)+
z\left(<\lambda^2 f>+ \right. \\ \left. <f><\lambda f>+\frac{1}{6}<f>^3\right),               \\
\theta_z=-\theta_x<f>-y\left(<\lambda f>-\frac{1}{2}<f>^2 \right)+
t\left(<\lambda^2 f>- \right. \\ \left. <f><\lambda f>+\frac{1}{6}<f>^3\right),
\ea
\eeq  
and manipulating equations (\ref{coeff2}) and (\ref{implicit_heav}), one shows that the 
solutions of (\ref{heavenly}) generated by the spectral data (\ref{manak_R}) satisfy the 
following linear PDE 
\beq\label{lin_const}
t\theta_t-z\theta_z+y\theta_y-x\theta_x=0.
\eeq 
Substituting in (\ref{heavenly}) the expression of $\theta_t$ given in (\ref{lin_const}), 
one obtains the lower dimensional nonlinear constraint
\beq
(x\theta_x)_x-y\theta_{xy}+z\theta_{xz}-t\theta_{yz}+t(\theta_{xx}\theta_{yy}-\theta^2_{xy})=0
\eeq 
satisfied by the solutions of (\ref{heavenly}) generated by the spectral data (\ref{manak_R}). 
  
%%%%%%%%%%%%%%%%%%%%%%%%%%%%%%%%%%%%%%%%%%%%%%%%%%%%%%%%%%%%%%%%%%%%%%%%%%%%%%%%%%%%%%%%%%%%%
%%%%%%%%%%%%%%%%%%%%%%%%%%%%%%%%%%%%%%%%%%%%%%%%%%%%%%%%%%%%%%%%%%%%%%%%%%%%%%%%%%%%%%%%%%%%%
\vskip 5pt
\noindent
{\bf Acknowledgements}. This research has been supported by the RFBR 
grants 07-01-00446, 06-01-90840, and 06-01-92053, by the bilateral agreement 
between the Consortium Einstein and the RFBR, by the bilateral agreement between the 
University of Roma ``La Sapienza'', by the Landau Institute for Theoretical Physics of the 
Russian Academy of Sciences and by a Visiting Professorship of the University of Roma ``La Sapienza''.  
%%%%%%%%%%%%%%%%%%%%%%%%%%%
%%%%%%%%%%%%%%%%%%%%%%%%%%%

\end{document}